\documentclass[fleqn,usenatbib]{mnras}
\usepackage{soul}
\usepackage{newtxtext,newtxmath}

\usepackage[T1]{fontenc}

\DeclareRobustCommand{\VAN}[3]{#2}
\let\VANthebibliography\thebibliography
\def\thebibliography{\DeclareRobustCommand{\VAN}[3]{##3}\VANthebibliography}

\usepackage{graphicx}	
\usepackage{amsmath}	

\usepackage{anyfontsize}

\usepackage{multirow}
\usepackage{multicol}
\usepackage{lipsum}
\usepackage{comment} 
\usepackage{makecell}

\usepackage{threeparttable}

\usepackage{subcaption}
\title[X-ray/gamma-ray correlation in HBL blazars]{On the correlation between X-rays and TeV gamma-rays in HBL Blazars.}

\author[M. Osorio-Archila et al.]{
M. Osorio-Archila,$^{1}$\thanks{E-mail: jmosorio@astro.unam.mx}
M. M. González,$^{1}$
J. R. Sacahui$^{2}$
\\
$^{1}$Instituto de Astronomía, Universidad Nacional Autónoma de México, Ciudad de México, México \\
$^{2}$IFIM, Escuela de Ciencias Físicas y Matemáticas,  Universidad de San Carlos de Guatemala, Guatemala, Guatemala
}

\pubyear{2024}

\begin{document}
\label{firstpage}
\pagerange{\pageref{firstpage}--\pageref{lastpage}}
\maketitle

\begin{abstract}


The gamma-ray emission in blazars can be attributed to the leptonic Synchrotron Self-Compton (SSC) model, photo-hadronic interactions, or a combination thereof. While evidence supports both models, their specific contributions remain uncertain. One supportive piece of evidence for the SSC model is the correlation between synchrotron and SSC fluxes in some blazar's Spectral Energy Distribution (SED), indicating the relative contributions of leptonic and hadronic mechanisms. Observational studies of the HBL blazar Markarian 421 over several years, spanning TeV gamma rays and X-rays, have reported a linear correlation across various timescales, which breaks at the highest gamma-ray fluxes. Extending this analysis to four High synchrotron peaked BL Lac (HBL) blazars -- Markarian 501, 1ES 1959+650, PKS 2155-304 and 1ES 2344+514.-- we utilize multiwavelength data from ground-based Imaging Atmospheric Cherenkov Telescopes (IACTs) for gamma rays and satellite observations for X-rays. Our long-term study confirms a linear correlation between fluxes across these energy bands, except for Markarian 501, which shows a correlation index of $1.45 \pm 0.01$. Notably, the exceptional flaring episode of PKS 2155-304 exhibits a correlation index of 2 with extreme values of gamma-ray fluxes. We observe outliers with high gamma-ray fluxes, suggesting the involvement of another mechanism, either of hadronic or leptonic origin. Finally, all other correlations exhibit alignment with a general correlation, suggesting a common acceleration mechanism among them with slight variations likely due to individual magnetic field strengths.

\end{abstract}



\section{Introduction}

BL Lacertae (BL Lac) objects are a subclass of blazars, which are a type of Active Galactic Nuclei (AGN) \citep{Padovani2017A&ARv..25....2P}. 
AGN derive their energy from the accretion of mass onto supermassive black holes (SMBH) and release it in the form of jets \citep{2019ApJ...875L...1E}.  For blazars, one of these jets is aligned, or nearly aligned, with the observer's line of sight from Earth \citep{Blandford1978bllo.conf..328B, 1995PASP..107..803U}. Blazars are known for their emission of radiation across the entire electromagnetic spectrum, with their Spectral Energy Distribution (SED) in the $\nu F_{\nu}$ vs. $\nu$ space exhibiting two distinct energy components at different  energy ranges. 
The SED classification of BL Lacs is based on the peak energy of the low-energy component, which categorizes them into three types: Low Synchrotron Peaked BL Lac (LBL), Intermediate Synchrotron peaked BL Lac (IBL) and High Synchrotron peaked BL Lac (HBL) \citep{PadovaniGiommi1995ApJ...444..567P}. In the case of HBL blazars, the peaks of the components, and hence most of their emission, are located in the X-ray and TeV gamma-ray bands. These HBL blazars are of particular interest because they constitute the majority of the extragalactic population of very high-energy gamma-ray emitters (VHE > 100 GeV).

In general, for all the spectral types, the low-energy emission component in these sources is well understood and is attributed to the synchrotron mechanism \citep{Blandford1978bllo.conf..328B, Blandford1979ApJ...232...34B}. 
The exact mechanism responsible for the high-energy emission component, however, is still not fully understood. This component cannot be solely explained by leptonic contributions, such as Synchrotron Self-Compton (SSC) or External Compton (EC) emission \citep{Wang2024ApJS..271...10W} with seed photons from sources like the accretion disk, Broad Line Region (BLR), dusty torus, Extragalactic Background Light (EBL), or Cosmic Microwave Background (CMB). 
The detection of uncorrelated variability and orphan flares, defined as an increase in gamma-ray flux without a corresponding increase in low-energy bands (e.g. \citet{Wang2022PhRvD.105b3005W}, and reference therein), along with limited but suggestive evidence of a connection between AGN and neutrinos observed by IceCube \citep{icecube2022evidence}, and the compelling extragalactic origin of Ultra-High-Energy Cosmic Rays (UHECR) \citep{abreu2022arrival}, provide strong indications for the inclusion of models like leptonic models with multiple emission zones and/or of hadronic models. Hadronic models propose that the emission of gamma-ray photons in AGN jets is attributed to cascades initiated by high-energy protons \citep{2014PhRvD..90b3007M}, requiring a significant presence of hadrons within AGN jets and an efficient acceleration mechanism. The specific circumstances of AGN that govern the relative contributions of different mechanisms to the TeV emission remains unresolved. \\
Under the SSC model, a correlation between the fluxes of the synchrotron and SSC emissions is expected \citep{katarzynski2010correlation}. This correlation results from the inherent interdependence between these emissions, with the underlying particle population and magnetic field strength. Any changes in these quantities will impact both emission components proportionally. However, in contrast to the SSC model, photo-hadronic models predict the emission of TeV photons independently from the low-energy emission. Therefore, photo-hadronic models do not anticipate a correlation between the TeV emission and the synchrotron emission. However, some authors have proposed the existence of a correlation between the emissions of protons and leptons when both particles are accelerated within the same region. In this scenario, the fluxes of these particles are interdependent due to the influence of the magnetic field on the acceleration of both particle types \citep{sol2013active}.

Therefore, the presence and strength of the correlation between low and high-energy fluxes may depend on the relative contributions of different emission mechanisms, and subsequently, it can be influenced by factors such as the number of emission regions or the energy regime, whether it is in the Thompson or Klein-Nishina (KN) regime \citep{tavecchio1998constraints}. For example, in cases where photo-hadronic processes play a significant role in the high-energy emission, the correlation between the fluxes may be weakened. The dominance of different mechanisms can vary among blazars, leading to diverse correlations. Hence, investigating the existence and robustness of the correlation can offer valuable insights into quantifying the relative contributions of both mechanisms and understanding the factors that determine which mechanism predominates the TeV emission in different blazars.

Correlations with diverse morphologies (linear, quadratic, and between both) have been reported in previous studies \citep{gliozzi2006long, katarzynski2005correlation, krawczynski2004multiwavelength, aharonian2009simultaneous_flare}. These correlations exhibit variations depending on factors such as the specific blazar being studied, the activity state, the observational campaign and the considered time scale. Nevertheless, the presence of these diverse morphologies emphasizes the need for comprehensive analysis that consider multiple variables and conditions. Some authors have examined the morphology of the flux correlation under specific considerations, in particular \citet{katarzynski2010correlation} proposed that the presence of different correlation indices can be explained by considering a jet with multiple emission zones. They suggested that the overall correlation index can be obtained by summing the individual correlation indices of each emission zone within the jet. Additionally, \citet{katarzynski2005correlation} suggested that in order to account for the observed linear correlation between X-rays and TeV gamma-rays, blazars must emit in the Klein-Nishina (KN) energy regime. In this regime, the interaction cross-section of high-energy photons in the inverse Compton (IC) process declines \citep{tavecchio1998constraints}.

The study conducted by \citet{gonzalez2019reconcilement} performed a comprehensive correlation analysis of the HBL blazar Mrk 421. The research utilized X-ray data in the energy range of 2-10 keV and gamma-ray data with energies greater than 400 GeV. The dataset covered the years from 1992 to 2009 and included both low and high activity states of the blazar. The study observed a strong linear correlation at different time scales, however, this correlation breaks down for high values of gamma-ray fluxes. Their results suggest that the previously perceived lack of correlation (e.g \citet{acciari2009ApJ...703..169A} for blazar Mrk 421) at low fluxes could be attributed to the limited coverage of flux ranges in those studies, emphasizing the importance of conducting long-term observations that encompass a wider range of fluxes. 

This paper presents a study of the correlation between X-ray and gamma-ray emissions from five extensively observed HBL blazars with redshifts z < 0.15: Mrk 501, 1ES 1959+650, PKS 2155-304 (including its remarkable flare observed in 2006), 1ES 2344+514 and Mrk 421. Public gamma-ray data from Imaging Atmospheric Cherenkov Telescopes (IACTs) and X-ray data from satellite observatories are required. The paper is structured as follows: Section \ref{sec:sample} provides the selection and description of the blazar sample, including details about the data used in the analysis. In Section \ref{sec:methodology}, we present the methods used to unify the datasets and determine the correlation model. Section \ref{sec:results} presents the results and discussion, and finally Section \ref{sec:conclusions} summarizes our findings.


\section{Sample selection}\label{sec:sample}

For our study, we select BL Lac blazars of HBL spectral type  based on the following criteria: (1) detection threshold above 200 GeV, and (2) reported redshift below 0.15, ensuring that the attenuation due to extragalactic background light (EBL) is below 20\% at energies around 300 GeV according to the \citet{franceschini2017extragalactic} EBL model.
From the TeVCat catalog\footnote{\label{TeVCat}\texttt{http://tevcat.uchicago.edu/} }, 18 sources satisfy our criteria. However, quasi-simultaneous X-ray and gamma-ray data were available for only five of these sources: Mrk 501, 1ES 1959+650, PKS 2155-304, 1ES 2344+514, and Mrk 421.

To facilitate comparison across all datasets, we standardized them to uniform units and energy thresholds for both X-ray and gamma-rays fluxes. Table \ref{tab:summary} summarizes the multi-frequency studies that contributed to our results. In the following subsections, we provide a concise overview of each individual source.

\subsection{Markarian 501}
\label{subsec:Mrk51}

Markarian 501 (Mrk 501) is a HBL blazar 
with Right Ascension (RA) of 16$^h$53$^m$52.2$^s$ and Declination (Dec) of +39$^{\circ}$45$''$37$'$ (J2000 equatorial coordinates obtained from TeVCat catalog, see footnote \ref{TeVCat}). With a redshift of z = 0.0337 \citep{Acciari11}, it was first detected in TeV gamma-rays together with Mrk 421 in 1996 by the Whipple Observatory \citep{quinn1996detection}. Mrk 501 exhibits characteristics of an Extreme High-Frequency Peaked Blazar (EHBL), with the synchrotron peak shifting up to two orders of magnitude \citep{krawczynski1999x, ahnenextreme, gliozzi2006long} during high-activity states \citep{Djannati1999A&A...350...17D, gall2009arXiv0912.4728G}. In contrast, the TeV peak increases only a few units of energy \citep{Acciari11}. Gamma-ray spectra from Mrk 501 in the 1-10 TeV energy range typically follow a Power Law (PL) function, nevertheless, the description of the spectra needs to take into account a curvature at lower energy ranges (e.g. \citet{samuelson1998tev, aharonian2001tev, furniss2015first}) and across different activity states. Correlations between X-ray and gamma-ray emissions have been observed over six observational periods, ranging from weak during low-activity states \citep{aleksic2015multiwavelength} to strong during periods of high activity \citep{gliozzi2006long, ahnenextreme}. The SED of Mrk 501 is predominantly modeled using a SSC scenario, although extreme parameter values are sometimes required to describe its high-energy emissions \citep{furniss2015first, cologna2017exceptional}. Multiple emission zone models have also been proposed to account for the synchrotron peak shift of two orders of magnitude observed during high activity states \citep{albert2007variable, aleksic2015multiwavelength, ahnenextreme}. Notably, Mrk 501 has exhibited orphan flares characterized by harder spectra, suggesting the possible involvement of hadronic processes \citep{Neronov2012A&A...541A..31N, Bartoli2012ApJ...758....2B}. In this study, we utilize datasets from six observational periods to investigate the correlation between X-rays and gamma-rays (see Table \ref{tab:summary}).

\subsection{1ES 1959+650}\label{sec:1ES1959+650}

The blazar 1ES 1959+650 
with RA of 19$^h$ 59$^m$ 59.8$^{s}$ and Dec of +65$^{\circ}$08$''$55$'$. It has a reported redshift of z = 0.048 \citep{Perlman96} and was first detected in X-rays in 1992 as part of the BL Lac object search program during the Einstein Slew survey \citep{Perlman96}. Subsequent observations in 1998 with the Utah Seven Telescope Array reported TeV gamma-rays from this source with a significance of 3.9$\sigma$ \citep{Nishiyama99}.

One distinctive characteristic of this blazar is the occurrence of gamma-ray orphan flares, with the first documented instance in June 2002 \citep{krawczynski2004multiwavelength}. The primary model used to explain its overall emission is typically the one-zone SSC model. However, extensive multi-wavelength flux correlations observed during various campaigns have indicated that this model cannot fully describe the SED \citep{krawczynski2004multiwavelength, santander2017exceptional, kapanadze2016long}. Moreover, the detection of orphan flares has prompted the exploration of additional theoretical frameworks, including the multizone SSC model \citep{kapanadze2018second, Sahu2021ApJ...906...91S}, the EC leptonic model, and various hadronic models \citep{krawczynski2004multiwavelength, Reimer05, Bottacini2010ApJ...719L.162B, santander2017exceptional}.

The position of the low-energy peak in this blazar shifts depending on its activity state. X-ray spectra are modeled using either a PL or a log-parabola function, depending on the location of the peak energy, often showing Compton dominance in the highest activity states \citep{Tagliaferri2008ApJ...679.1029T}. 
The high-energy component typically follows a PL function for energies above 300 GeVs. The high-energy peak has been observed ranging from 0.039 TeV in the lowest activity state \citep{uellenbeck2013study} to between 0.4 and 0.7 TeV in some of its most active states \citep{acciari2020broadband}. The present study uses data from five multiwavelength campaigns, detailed in Table \ref{tab:summary}, to analyze the correlation between X-rays and TeV gamma-rays.

\subsection{PKS 2155-304}

The blazar PKS 2155-304 
with RA of 21$^{h}$ 58$^{m}$ 52.7$^{s}$ and Dec of -30$^{\circ}$13$''$18$'$. This source has a reported redshift of z = 0.117 \citep{Aharonian05}. At this redshift, the flux observed at 200 GeV is attenuated by approximately 11$\%$ due to the EBL \citep{franceschini2017extragalactic}. 

PKS 2155-304 was first observed in X-rays in 1979 by the HEAO-1 satellite \citep{schwartz1979x} and detected at TeV energies in 1996 by the Durham Mark 6 Cherenkov observatory \citep{chadwick1999very}. Since 2004, it has been regularly monitored by the High Energy Stereoscopic System (HESS) \citep{abdalla2017characterizing}. A long-term study by \citet{gliozzi2006long}, which did not take into account an exceptional flare observed in 2006 July 29, identified a correlation between X-rays and gamma-rays. During the exceptional flare, the gamma-ray fluxes increased by up to two orders of magnitude, and \citet{aharonian2009simultaneous_flare} reported a quadratic correlation when analysing data for the whole campaign. The occurrence of this exceptional flare, along with orphan flares observed in the optical and ultraviolet bands \citep{wierzcholska2019hess}, challenges the adequacy of the one-zone SSC model, suggesting a preference for the multiple-zone SSC model \citep{Aharonian2005A&A...442..895A, aharonian2007exceptional, Wang2022PhRvD.105b3005W} or a hadro-leptonic model \citep{abdalla2020simultaneous,aharonian2007exceptional}. The high-energy peak in the SED varies with the activity state, ranging from $\sim$20-50 GeV during periods of low activity \citep{Weidinger2010ASTRA...6....1W, aharonian2009simultaneous} and up to 500 GeV during periods of high activity \citep{aharonian2007exceptional}. The present study utilizes data from four observational periods spanning from 2006 to 2016 (see Table \ref{tab:summary}). None of the data were corrected for EBL.

\subsection{1ES 2344+514}

1ES 2344+514 is a BL Lac blazar 
with RA of 23$^{h}$ 47$^{m}$ 04$^{s}$ and Dec of +51$^{\circ}$42$''$49$'$. It has a reported redshift of z = 0.044, making it the third closest blazar after Mrk 421 and Mrk 501. Initially detected in X-rays with energies ranging from 0.2 to 4 keV by the Einstein Slew Survey, it was later observed in December of 1995 in TeV gamma-rays with energies exceeding 300 GeV by the Whipple 10m telescope \citep{catanese1998discovery, acciari2020intermittent}. During periods of high activity, the synchrotron peak of 1ES 2344+514 shifts to higher energies by a factor of 20-30 \citep{Schroedter2005ApJ...634..947S, albert2007observation}, classifying it as an Extreme High-Frequency Peaked Blazar (EHBL). Similarly, the high-energy peak shifts from 40 GeV to 400 GeV during these high active states \citep{acciari2011multiwavelength, acciari2020intermittent}. The SED of both low and high states has been successfully modeled using the one-zone SSC model \citep{albert2007observation}. Alternative models such as a two-zone emission model \citep{2013A&A...556A..67A, MAGIC2020A&A...640A.132M} and more recently, a two-zone proton-synchrotron model \citep{Wang2024ApJS..271...10W}, have been tested with promising results.

\subsection{Markarian 421}

Markarian 421, or Mrk 421 
with RA of 11$^{h}$ 04$^{m}$ 19$^{s}$ and Dec of +38$^{\circ}$11$''$41$'$. It is notable for being the closest blazar to Earth, with a redshift of z = 0.034, and was the first blazar detected in gamma-rays, in 1992, by the Whipple Observatory \citep{punch1992detection}. Mrk 421 together with  Mrk 501 are the most luminous blazars in TeV energies, however they demonstrate intrinsic differences. Specifically, in Mrk 421, both the low and high-energy components of its SED, are cut at lower energies compared to Mrk 501 \citep{Krennrich2001ApJ...560L..45K}. This distinction suggests potential differences in particle diffusion mechanisms, photon absorption characteristics, or fundamental physical parameters such as magnetic field or the maximum energy of electrons \citep{Baheeja2024PhRvD.109j3039B, Aharonian1999A&A...350..757A, korochkin2021sensitivity}. During periods of high activity, both Mrk 421 and Mrk 501 exhibit a similar increase in their SED flux, along with a shift towards higher energies, up to two orders of magnitude in their synchrotron peak, although at different energy ranges. The low-energy peak of Mrk 421 varies from 0.1 to 10 keV \citep{Tramacere2007A&A...466..521T, Kapanadze2017ApJ...848..103K}, whereas in Mrk 501, this shift can reach up to 200 keV. Furthermore, when comparing their high-energy peak, Mrk 421 presents a higher flux \citep{2007ApJ...663..125A}. The SED of Mrk 421 is typically modeled using the one-zone SSC leptonic model \citep{Bloom1993AIPC..280..578B, Wang2024ApJS..271...10W}. However, due to the observed shifts in the SED components and the occasional lack of correlation in variability across different energy bands, alternative models such as a two-zone SSC scenario \citep{blazejowski2005multiwavelength, abeysekara2020great, sahu2021extreme} or a hadronic model \citep{Wang2024ApJS..271...10W} have been proposed. In the present study, we take the analysis framework of the linear correlation reported by \citet{gonzalez2019reconcilement} based on 14 years of data \citep{Acciari2014APh....54....1A}.

\begin{table*}
    \centering
    \caption{
    Important information from each multiwavelength study used for our sample. 
    }
    \label{tab:summary}
\resizebox{\textwidth}{!}{
\begin{tabular}{lccccccccccc}

\hline
\hline
Period & \multicolumn{2}{c}{Energy$^{a}$} & TS$^{b}$ & Avg Sim$^{c}$ & SA$^{d}$  &  $\rm E_{p, TeV}{ }^{e}$ & Spectral$^{f}$ & Spectral$^{g}$& \multicolumn{2}{c}{Instruments$^{h}$} & Ref \\ 
yy mm - yy mm & (keV) & (TeV) & & (hr) & & (TeV) & model & index & X-rays & gamma-rays & \\ \hline \hline

\multicolumn{11}{c}{Markarian 501}   \\ \hline
1) 1997 Mar - 1997 June  & $2-20$ & \textgreater 1   & N & 2.25    & A      & 1          & PL   & 2.25 & RXTE-PCA & HEGRA  & (1, 2, 3) \\
2) 1998 Feb - 1998 July  & $2-20$ & \textgreater 1   & N & 2.41    & Q     & 0.2        & PLCO & 2.31 & RXTE-PCA & HEGRA  & (1, 4, 3) \\
3) 1999 May - 1999 July  & $2-20$ & \textgreater 1   & N & 7.57    & Q       & 0.2        & PLCO & 2.31 & RXTE-PCA & HEGRA  & (1, 4, 3) \\
4) 2008 Mar - 2008 May  & $2-10$ & \textgreater 0.3 & N & 1.51    & Q     & 0.04       & PL & 2.42 \& 2.47 & Swift-BAT & M \& V  & (5)\\
5) 2012 Mar - 2012 July  & $2-10$ & \textgreater 1   & N & 2.86    & A     & $\sim$ 1   & PL & 2.08 \& 1.88 & Swift-XRT & M \& V & (6) \\
6) 2013 Apr - 2013 Aug  & $2-10$& 
\textgreater 0.2 & - & 4.7     & Q     & $\sim$ 0.04 & PL & 2.29 \& 2.37 & Swift-XRT & M \& V  & (7)\\ \hline \hline
\multicolumn{11}{c}{1ES1959+650}   \\ \hline 
1) 2002 May - 2002 Aug  & 10$^{\dag}$     & \textgreater 0.6 & - & 0.5    & A/Q       & 0.3         & PL & 2.80 & RXTE-PCA  & Whipple  & (8, 9)       \\
\ \ \ \ \ 2002 May - 2002 Aug  & 10$^{\dag}$  & \textgreater 2   & - & 1.6    & Q         & 0.3         & PL & 3.18 & RXTE-PCA  & HEGRA   & (8, 10)       \\
2) 2009 June - 2009 Sept  & $0.3-10$&\textgreater 0.3 & D & 7.17   & Q        & 0.035       & PL & 2.56 & Swift-XRT & MAGIC    & (11$^{*}$, 12) \\
3) 2012 Apr - 2012 June  & $2-10$ &\textgreater 0.315& - & 2.97   & A         & 0.09        & PL & 2.60 & Swift-XRT & VERITAS   & (13, 14$^{*}$) \\
4) 2015 Oct - 2016 June  & $0.3-10$&\textgreater 0.3 & N & 3.72   & A/Q    & 0.1         & PL & 2.77 & Swift-XRT & VERITAS  & (15, 16, 17$^{*}$)     \\
5) 2016 May - 2016 Nov  & $0.5-5$& \textgreater 0.3 & N & 6.21   & A/Q    & 0.7         & PL & 2.06 & Swift-XRT & MAGIC    & (18, 14)     \\ \hline \hline
\multicolumn{11}{c}{PKS2155-304}   \\ \hline   
1) 2006 Aug - 2012 Oct & $0.3-10$ & \textgreater 0.3& N & 8.09    & A/Q       & -           & PL & 3.53 & Swift-XRT& HESS  & (19, 20)\\
  \ \ \ \ \ 2006 July 29      & $0.5-5$ & \textgreater 0.2 & 8m& 0.03 & EH   & 0.5         & PLCO & 2.86 & Chandra-LETG & HESS   & (21, 22) \\
2) 2008 Aug - 2008 Sept & $2-10$ & \textgreater 0.2  & N & Total  & A/Q       & 0.01        & PL & 3.34 & Swift-XRT & HESS  & (23) \\
3) 2013 June - 2013 Sept & $2-10$ & Several$^{\dag \dag}$  & N & 7.39   & Q         & 0.01        & PL & 3.00 & Swift-XRT&HESS  & (24) \\
4) 2015 May - 2016 Aug & $0.3-10$&\textgreater 0.2  & D  & 2.87  & Q        & -           & - & - & Swift-XRT&HESS    & (25, 14$^{*}$) \\ \hline \hline 
\multicolumn{11}{c}{1ES 2344+514}   \\ \hline   
1) 2007 Oct - 2008 Jan & $2-10$ & \textgreater 0.3  & N & 1.05   & A/Q    & 0.5         & PL & 2.61 & \makecell{ RXTE-PCA \&  \\ Swift-XRT } & VERITAS  & (26) \\ \hline \hline
\multicolumn{11}{c}{Markarian 421}   \\ \hline   
1) 1992 - 2004 & $2-10$ & \textgreater 0.4  & M &  -  & A/Q   &  -    & PLCO & 2.11-2.75 & RXTE-PCA & Whipple   & (27, 28, 29, 30) \\ \hline \hline
\end{tabular}
}
    \begin{tablenotes}
    \item Notes: 
    \item $(a)$ The energy threshold originally reported for X-rays and gamma-rays for each period, respectively. $(b)$ The time scale (TS) for each data set: (N) denotes nightly time scale, (D) denotes daily time scale, and (8m) denotes an 8-minutes time scale. $(c)$ The average time simultaneity between X-rays and gamma-rays for each data set. $(d)$ The state of activity (SA) observed in each study: (A) for active state, (Q) for quiescent state, or (A/Q) for a combination of both; (EH) denotes the extremely high flare reported in PKS 2155-304 in 2006. $(e)$ The maximum TeV energy peak reported in each work. $(f)$ Gamma-ray spectral model reported for each period of observation, (PL for Power Law and PLCO for Power Law with exponential Cut Off), following the average photon spectral index $(g)$. $(h)$ The instruments that conducted the observation. $\dag$ In this case, the value reported is the dN/dE at 10 keV of energy. $\dag\dag$ Each flux point was standardized to the same energy threshold using the parameters reported in (24). $^*$ The Swift-XRT X-ray data used the conversion factor reported in \citet{stroh2013swift}, assuming the conversion factors did not undergo a significant variation for the datasets belonging to years above 2012. 
    \item \textsc{References}.--- 
    (1) \citet{gliozzi2006long}, 
    (2) \citet{aharonian1999temporal} (see \citep{hillas1998spectrum} for conversion factor from CU to cgs photon flux), 
    (3) RXTE database: (\texttt{https://cass.ucsd.edu/$\sim$rxteagn/Mkn501/Mkn501.html}), 
    (4) \citet{aharonian2001tev},
    (5) \citet{aleksic2015multiwavelength},
    (6) \citet{ahnenextreme},
    (7) \citet{furniss2015first},
    (8) \citet{krawczynski2004multiwavelength}, 
    (9) \citet{daniel2005spectrum} (see \citep{hillas1998spectrum}), 
    (10) \citet{aharonian2003detection},          
    (11) \citet{kapanadze2016long},
    (12) \citet{uellenbeck2013study},
    (13) \citet{aliu2014investigating}, 
    (14) Swift-XRT database \texttt{https://www.swift.psu.edu/monitoring/},
    (15) \citet{santander2017exceptional}, 
    (16) \citet{kapanadze2016recent}, 
    (17) \citet{kapanadze2018second},  
    (18) \citet{acciari2020broadband}, 
    (19) \citet{goyal2020blazar}, 
    (20) \citet{abdalla2017characterizing}, 
    (21) \citet{aharonian2009simultaneous_flare}, 
    (22) \citet{abramowski2010vhe}, 
    (23) \citet{aharonian2009simultaneous},
    (24) \citet{abdalla2020simultaneous},
    (25) \citet{wierzcholska2019hess}, 
    (26) \citet{acciari2011multiwavelength},
    (27) \citep{Acciari2014APh....54....1A}, 
    (28) \citet{Grube2008ICRC....2..691G}, 
    (29) \citet{Accicari2011ApJ...738...25A}
    (30) \citet{smith2002x}.
    \end{tablenotes}
\end{table*}


\section{Methodology} \label{sec:methodology}

\subsection{Data Standardization}
\label{sec:data_stand}

In Table \ref{tab:summary} we summarize the information for each source used in our analysis, including data from Mrk 421 from \cite{Acciari2014APh....54....1A}. 
The table displays multiple datasets for each source across various observational campaigns, each with different energy thresholds and flux units. It consists of 12 columns. The first column indicates the period of each dataset. Up to the 11th column the information belongs to how the datasets were originally reported. Column 12th contains the references from where we obtained the datasets, the spectral models used to modify the energy threshold, and the unit conversion factors when needed. To standardize these datasets, we calculate integral fluxes up to a previously specified energy threshold for gamma-rays and within defined energy intervals for X-rays per source, as detailed in Table \ref{tab:sources}, using the corresponding spectral models reported in the literature. It is important to note that the spectral model may vary between measurements within a single campaign or period, as indicated by the references in Table \ref{tab:summary}. When only an average spectral model is reported, it is applied uniformly across the dataset, potentially resulting in an over or underestimation of the calculated integral fluxes. We utilize the cgs unit system, using ph cm$^{-2}$ s$^{-1}$ for photon flux units for gamma-rays, and erg cm$^{-2}$ s$^{-1}$ for energy flux units for X-rays. For Mrk 421, gamma-rays fluxes were provided in Crab units, which we convert using the Crab flux values from  \cite{Grube2008ICRC....2..691G}. The RXTE-ASM X-ray count rates are converted to energy flux using a conversion factor of 1 cts s$^{-1}$ = 3.71$\times 10^{-10}$ erg cm$^{-2}$ s$^{-1}$ in the energy range from 2 to 10 keV, as reported by \cite{smith2002x}. For the Swift-XRT data, we use the conversion factors reported in \citet{stroh2013swift}. Since the remaining X-ray data were published using erg cm$^{-2}$ s$^{-1}$ units, they remained without change.

\begin{table*}
    \centering
    \caption{
    Gamma-ray energy threshold and X-ray energy range used for each source.
    }
    \label{tab:sources}
\begin{tabular}{ccc}

\hline
\hline
Source & $\rm E_{th,\gamma}$ (TeV) & $\rm \Delta E_{X}$ (keV) \\ \hline \hline 
Mrk 501       &   1 & 2-10   \\
1ES 1959+650  & 0.3 & 0.3-10 \\
PKS 2155-304  & 0.2 & 2-10   \\
1ES 2344+514  & 0.3 & 2-10   \\
Mrk 421       & 0.4 & 2-10   \\ \hline \hline 
\end{tabular}
\end{table*}

\subsection{Correlation models} \label{subsec:correlationmodels}

We consider a correlation model of the form $F_{\gamma} = bF_{X}^{\alpha}$, where $F_{\gamma}$ is the gamma-ray flux, $F_{X}^{\alpha}$ is the X-ray flux, $\alpha$ is the correlation index and $b$ is the normalization factor, and test three hypotheses: linear ($\alpha=1$), quadratic ($\alpha=2$), and a free value of the index $\alpha$. According to the SSC model, a quadratic dependency between X-ray and gamma-ray fluxes is expected \citep{amenomori2003multi} when there is variation only in the electron density \citep{singh2019multi}. This is often observed for single flares \citep{aharonian2009simultaneous_flare}. The SSC model does not predict a combination of terms of different orders such as linear, quadratic, cubic, etc. However, when the emission originates from multiple zones, as may be the case for our analysis including flaring episodes at different times, the value of $\alpha$ is expected to be between 1 and 2 \citep{katarzynski2010correlation}. Finally, $\alpha$ is expected to have a value of 1 when the emission occurs in the Klein-Nishina regime \citep{amenomori2003multi, katarzynski2005correlation}. This is the value found by \citet{gonzalez2019reconcilement} for Mrk 421.

\subsection{Statistical method} \label{sec:statistical}

To obtain the most accurate description of the correlation, we employ a Bayesian statistical method developed by \citet{d2005fits}. This approach utilizes Maximum likelihood estimation to fit a model to the data, accounting for inherent unknown data scattering ($\sigma_s$). This additional scatter is assumed to follow a normal distribution and is treated as a standard deviation. We interpret the points falling within 3$\sigma_s$ (99.7$\%$ of probability) as consistent with the correlation, while points outside this range -- referred to as outliers -- are considered deviations from the correlation.

For a PL model $F_{\gamma} = b F_{x}^{\alpha}$, the likelihood function ($L$) used to optimize the values of the free parameters $\omega$=\{$\alpha$, $b$\}, and  of the extra scatter of the data $\sigma_s$, is as follows:

\begin{equation}
\begin{split}
    L(\omega, \sigma_s; x, \gamma)= \frac{1}{2} \sum log[\sigma_s^2 + \sigma_{\gamma}^2 + F_{\gamma}^{\prime 2}(x, \omega) \sigma_{x}^2] \\ + \frac{1}{2} \sum \frac{[\gamma - F_{\gamma}]^2}{\sigma_s^2 + \sigma_{\gamma}^2 + F_{\gamma}^{\prime 2}(x, \omega) \sigma_{x}^2},
    \label{eq:DAgostini_PowerLaw}
\end{split}    
\end{equation}
where $x$ and $\gamma$ represent the measured X-ray and gamma-ray fluxes, respectively, with $\sigma_x$ and $\sigma_\gamma$ denoting their corresponding uncertainties.

To determine which correlation model best fits the datasets, we compare them using the Akaike Information Criterion (AIC) \citep{akaike1987factor} which is defined as follows for small sample sizes,

\begin{equation}
    \mathrm{AIC} = -2\ln(L) + 2w + \frac{2w(w+1)}{p-w-1},
\end{equation}
where $w$ is the number of free parameters, and $p$ is the sample size.
The AIC compares two models based on their likelihood values and adds penalty terms proportional to the number of free parameters. Therefore, if two models have similar likelihood, the AIC value will be higher for the model with more free parameters. A small difference in AIC values between two models may indicate that they are essentially equivalent. To assess the significance of the difference in the AIC values, we calculate the Relative Likelihood \citep{burnham2002model}, defined as,

\begin{equation}
    \mathrm{RL_{l/h} = exp\left(\frac{AIC_l - AIC_h}{2}\right) }
\end{equation}
where $\mathrm{AIC_l}$ has a lowest value than $\mathrm{AIC_h}$. The relative likelihood represents the probability that the model with $\mathrm{AIC_h}$ provides a good fit to the data, compared to the model corresponding to $\mathrm{AIC_l}$ (see Table \ref{tab:fitresult}).


\section{Results and Discussion}\label{sec:results}

In the study conducted by \citet{gonzalez2019reconcilement}, a correlation was identified for the blazar Mrk 421 over an extended period of time and across various flux states and time scales. In this study, we compile multiple datasets to investigate whether similar correlations exist in four other HBL blazars: Mrk 501, 1ES 1959+650, PKS 2155-304, and, 1ES 2344+514. To establish these correlations, we collect and standardize quasi-simultaneous X-ray and TeV gamma-ray observations within a few months and up to 16 years (refer to Table \ref{tab:summary}). In most of the observational campaigns the data was reported in daily time scales. The time scales of the light curves could introduce a bias, especially when sources exhibit flux variability in time scales less than a few hours. Our analysis and interpretation of the results are constrained by the average flux behavior within these time scales. This could be one source of the extra scatter of the data which is taken into account in the \citet{d2005fits} fit method. 

We also analyze the data corresponding to the exceptionally bright flare of PKS 2155-304 separately, which exhibits different behavior compared to other PKS 2155-304 datasets. We then assess the existence of a correlation by testing the models described in Section \ref{subsec:correlationmodels} and we use the Relative Likelihood method for model selection, as detailed in Section \ref{sec:statistical}. Table \ref{tab:aic} presents the correlation index when allowed to vary freely, along with the AIC values for each model, and their Relative Likelihood comparison. Our analysis reveals a correlation between X-ray and gamma-ray fluxes in all blazars studied. The best-fit model is highlighted in bold font. For four of the examined blazars, a linear correlation provides the most appropriate description of the relationship between fluxes. However, for the bright flare of PKS 2155-304, a quadratic model is preferred. In the case of Mrk 501, neither a linear nor quadratic model is preferred; instead, the correlation is characterized by an index value of $\alpha = 1.45 \pm 0.01$. 

For long-duration observations (i.e. longer than weeks), a linear correlation is typically expected \citep{gliozzi2006long, 2007ApJ...663..125A, gonzalez2019reconcilement}, as observed in all four blazars in our sample. A plausible explanation is that long-duration observations likely encompass emissions from multiple emission zones. These blazars are likely emitting in the Klein-Nishina regime, as discussed by \citet{katarzynski2005correlation}. Conversely, during short flaring periods, such as the one observed in PKS 2155-304, quadratic or cubic correlations are commonly reported \citep{Fossati08, aharonian2009simultaneous_flare}, which aligns with the finding of this study. 

\begin{figure*}
	\centering
	\begin{subfigure}{0.4\linewidth}
	     \includegraphics[width=\linewidth]{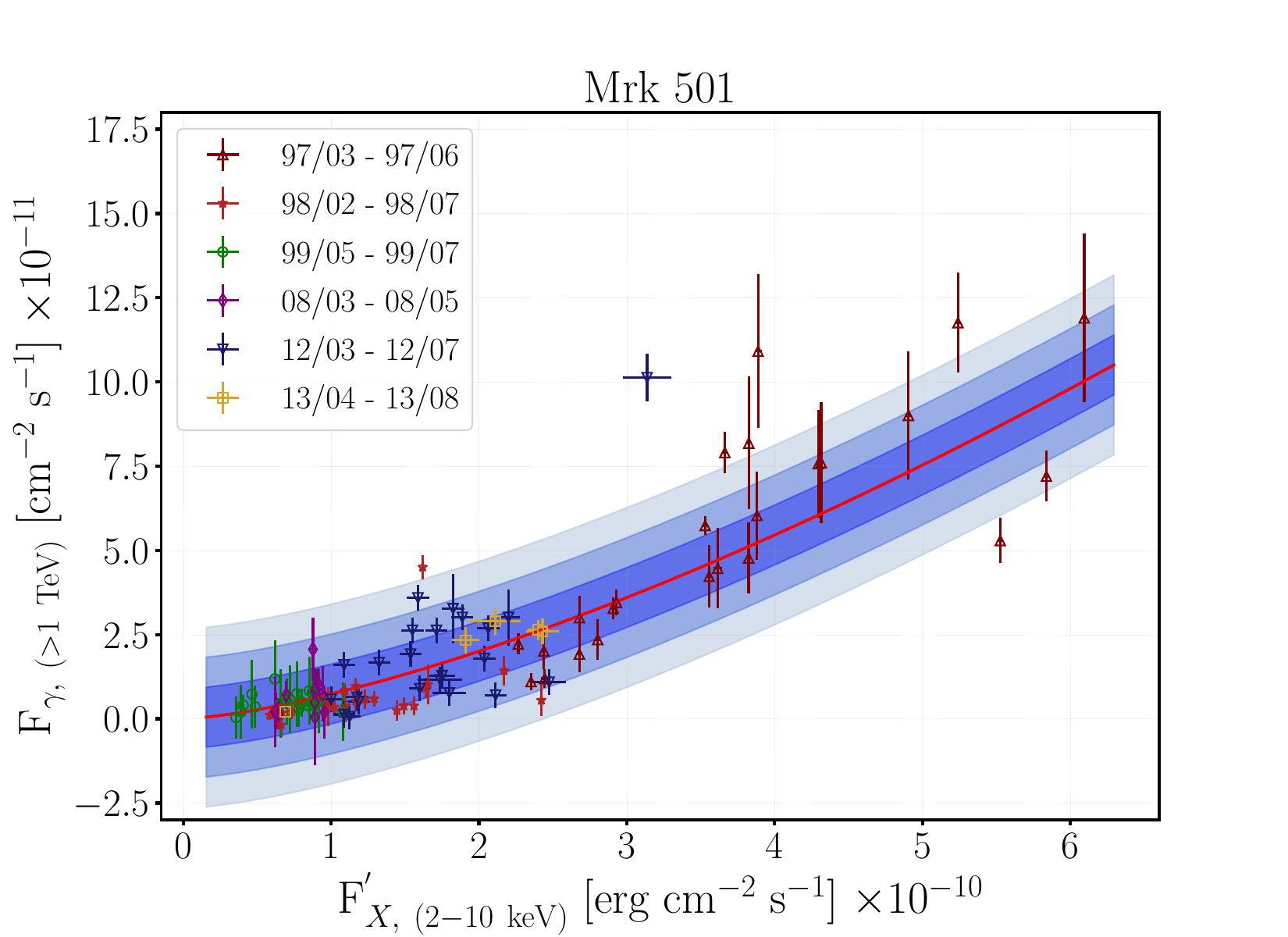}
      \caption{}
	\end{subfigure}
	\begin{subfigure}{0.4\linewidth}
	    \includegraphics[width=\linewidth]{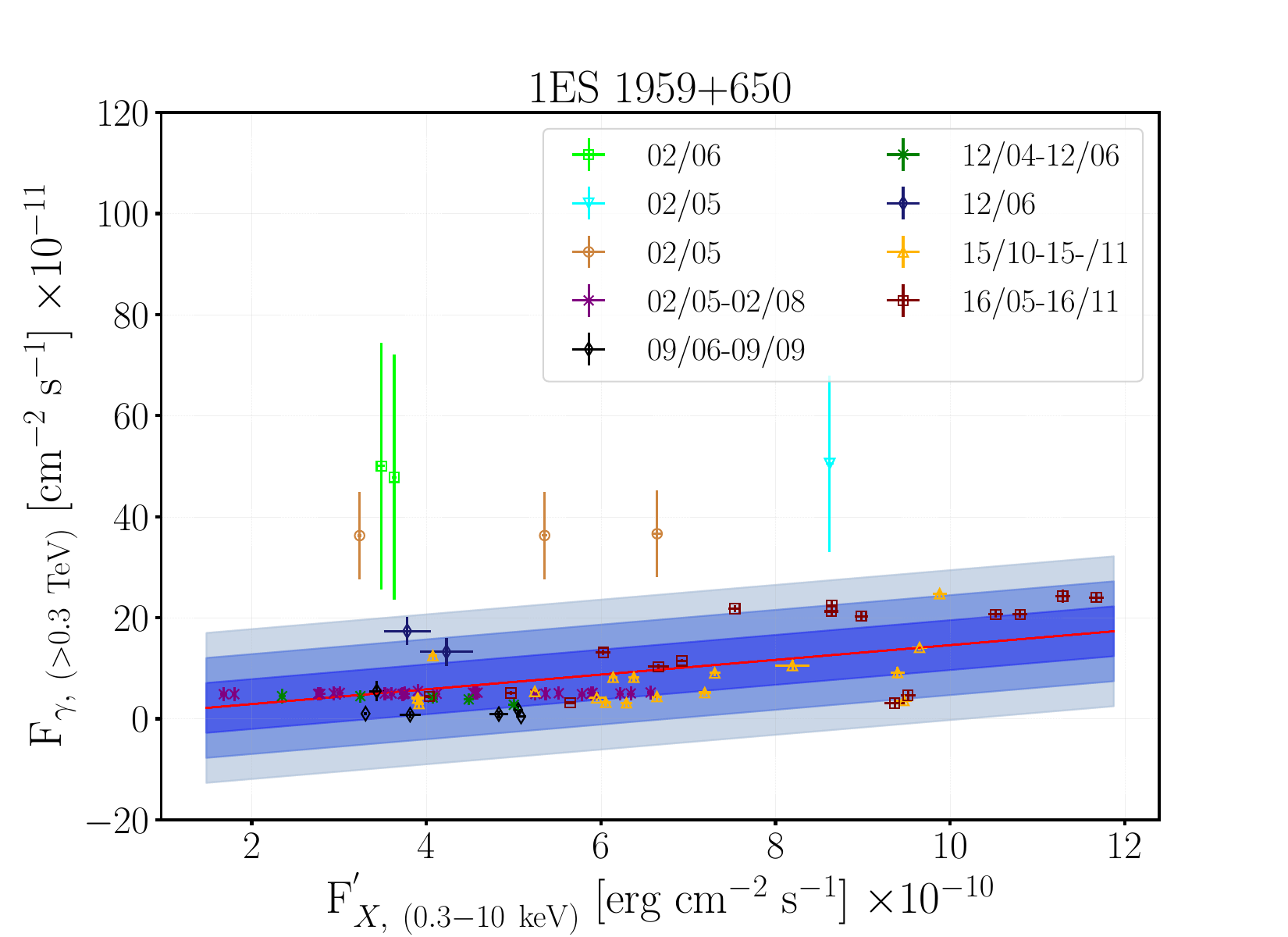}
        \caption{}
    \end{subfigure} \\	
	\begin{subfigure}{0.4\linewidth}
	    \includegraphics[width=\linewidth]{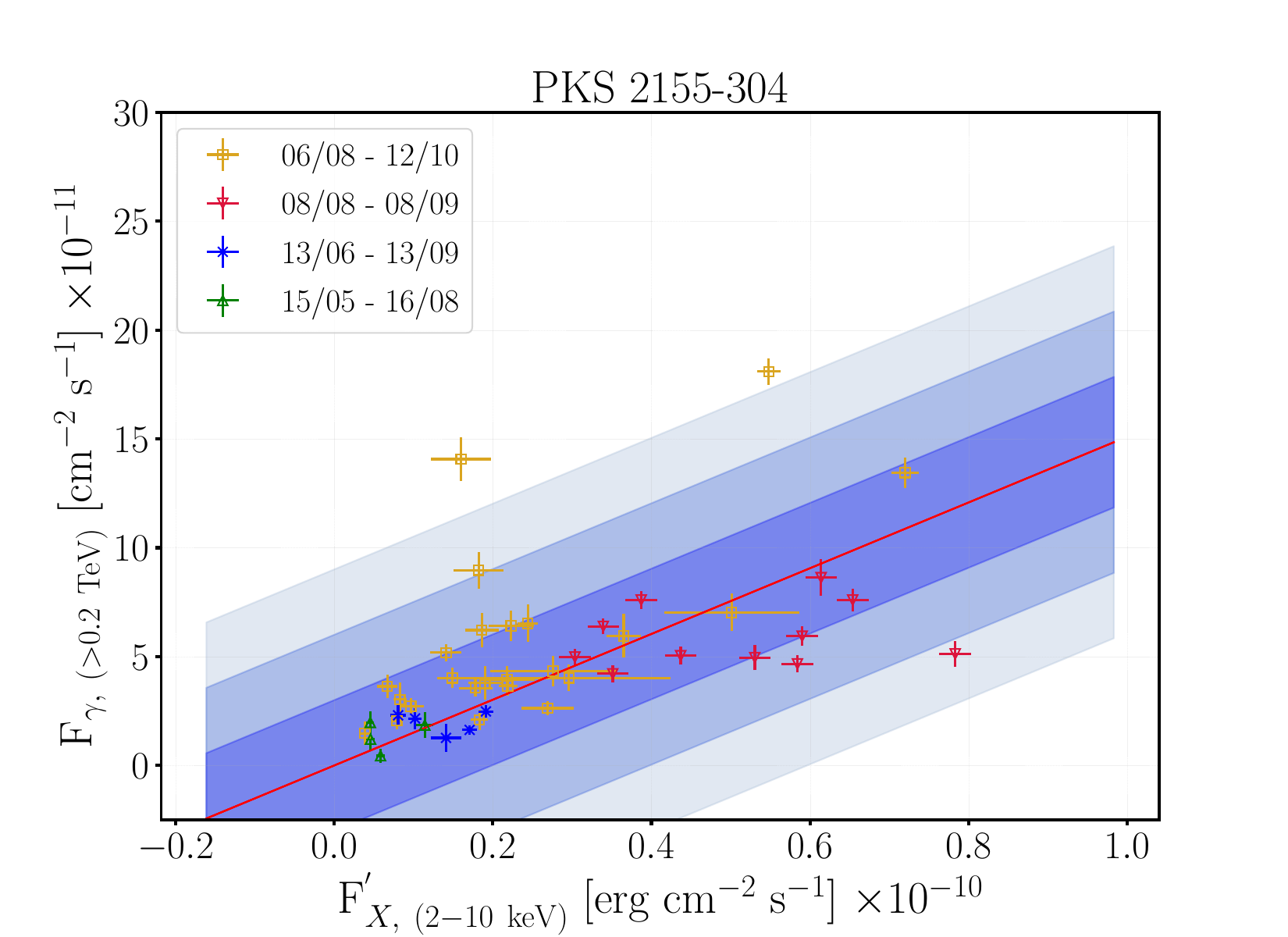}
        \caption{}
    \end{subfigure}
    \begin{subfigure}{0.4\linewidth}
	    \includegraphics[width=\linewidth]{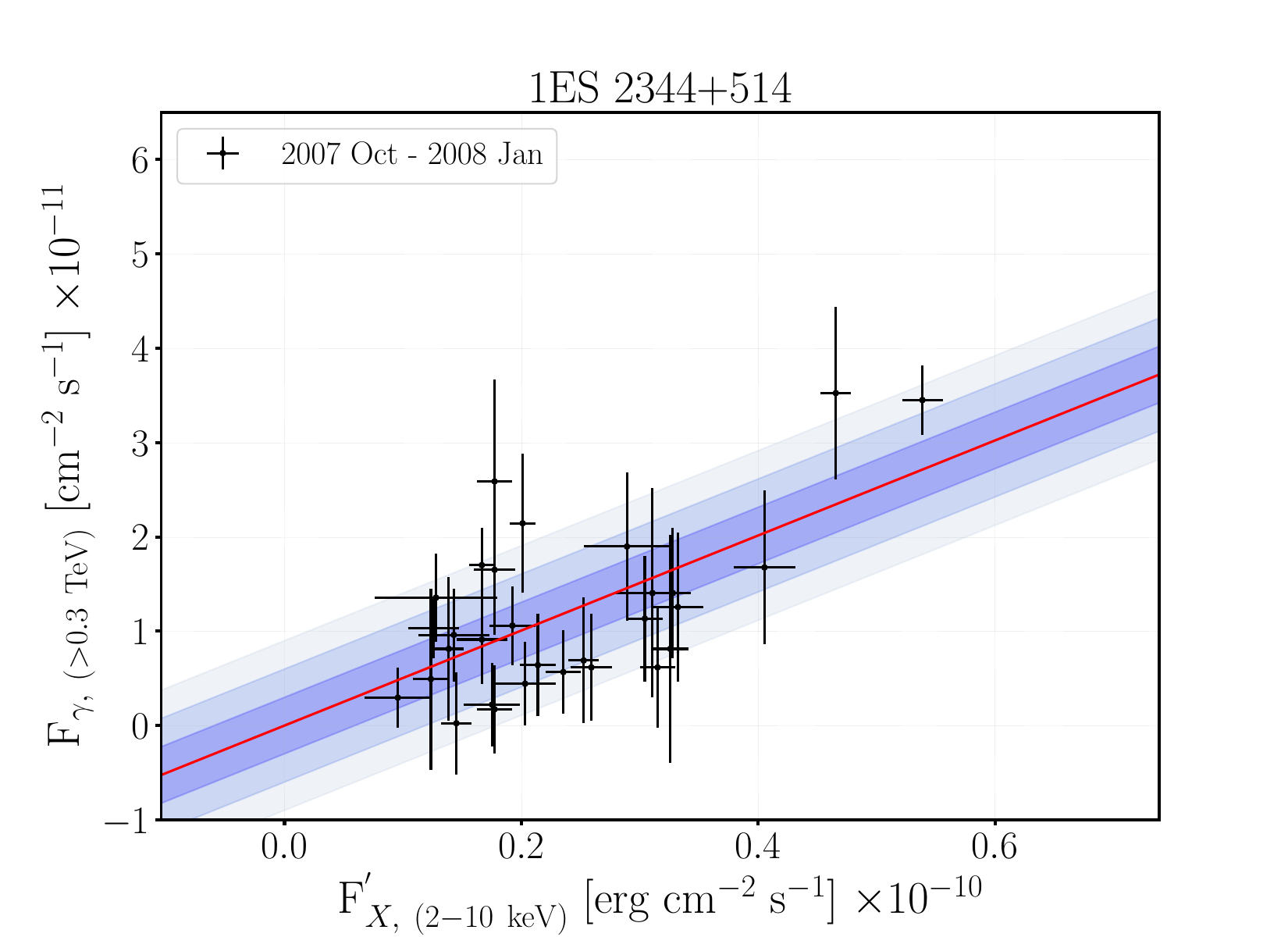}
        \caption{}
    \end{subfigure}
    \begin{subfigure}{0.4\linewidth}
	    \includegraphics[width=\linewidth]{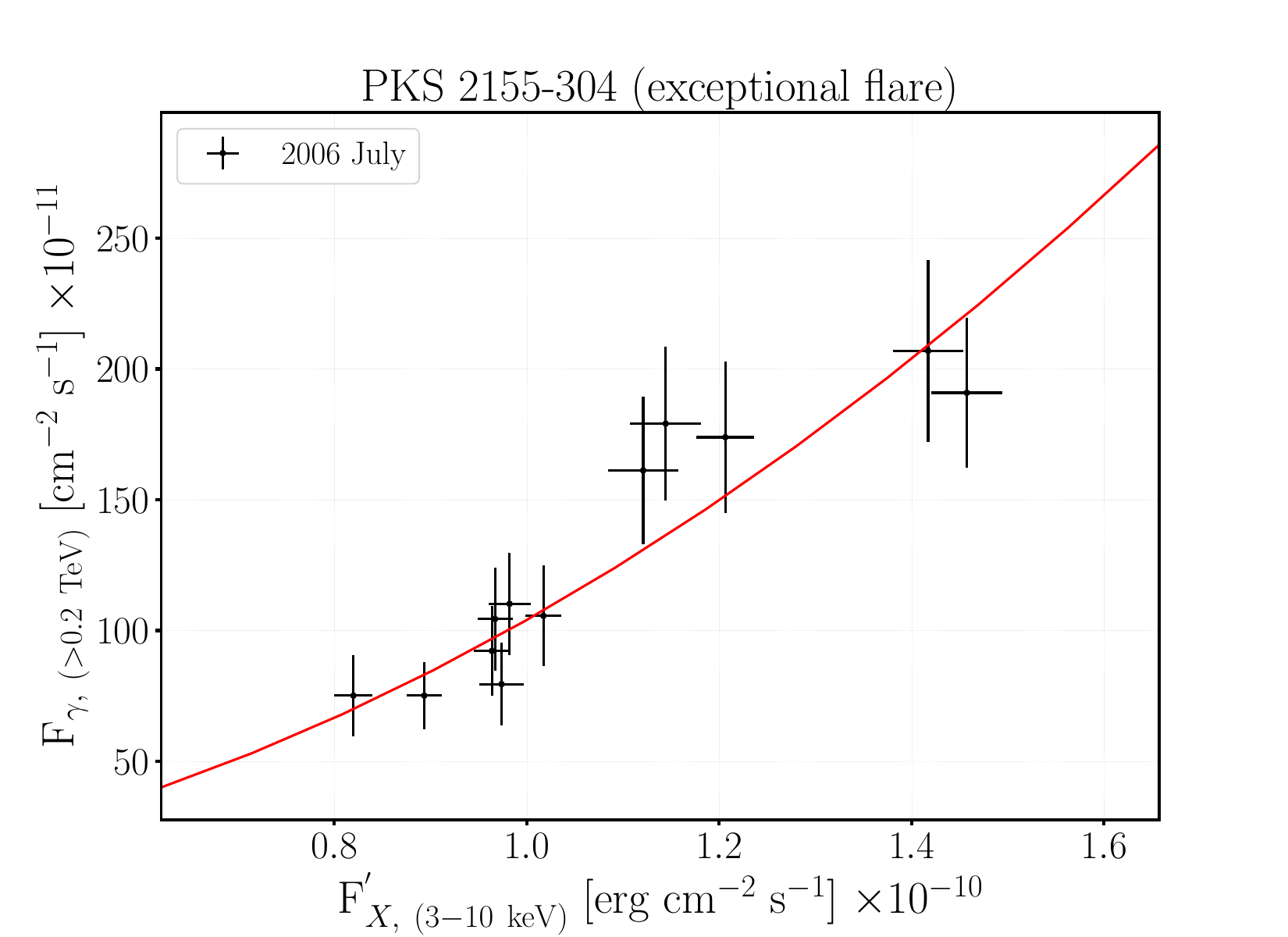}
        \caption{}
    \end{subfigure}
    \begin{subfigure}{0.4\linewidth}
	    \includegraphics[width=\linewidth]{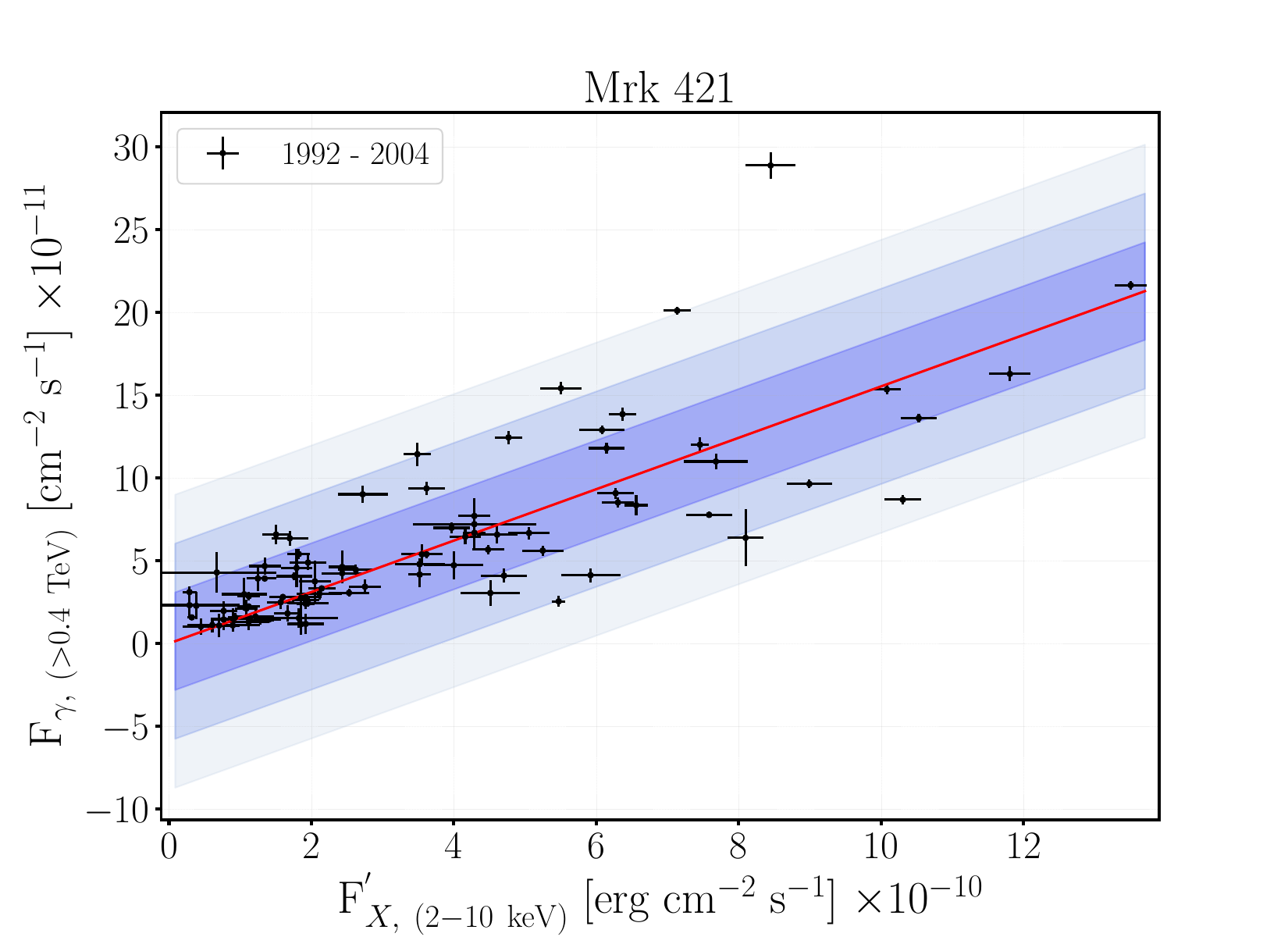}
        \caption{}
    \end{subfigure}

\caption{
Correlation between X-rays and gamma-rays for each blazar of our sample. The colors of the symbols in each panel correspond to different observation periods (see Table \ref{tab:summary}). The correlations are modeled using a PL with index fixed at $\alpha = 1$, except for Mrk 501 (panel a) and the exceptional flare of PKS 2155-304 (panel e), for which the PL index is free to vary, resulting on indices values of $\alpha = 1.45 \pm 0.09$ and $\alpha = 1.95 \pm 0.28$, respectively. The best model that describes the correlations is shown with the red solid line, the blue shaded regions correspond to 1, 2 and 3 times the $\sigma_s$. We observe outliers when analyzing correlations across multiple observational campaigns that consider high-activity gamma-ray fluxes. Labels indicate the year and month of the observational campaign.}
\label{fig:singlecorrelation}
\end{figure*}

In Figure \ref{fig:singlecorrelation} we present the best-fit model describing the correlation between X-ray and gamma-ray fluxes for each blazar, including the exceptionally bright flare of PKS 2155-304 (see panel e). The red solid lines represent the identified correlations, while the blue shaded regions indicate the permissible scatter defined by one, two, and three times $\sigma_s$. Notably, the value of $\sigma_s$ is significantly greater for PKS 2155-304 when the exceptionally bright flare is excluded, in contrast to the negligible value found during its very bright flare, which was monitored on a minute-time scale. Generally, the data aligns well with the identified correlations when considering an additional scatter up to 3$\sigma_s$. However, except for 1ES 2344+514, where only one multi-wavelength study is available, a few outliers are observed in all blazars, particularly at the highest gamma-ray fluxes for each blazar, as previously noted for Mrk 421 by \cite{gonzalez2019reconcilement}. 

Interestingly, only one outlier, depicted in light green for 1ES 1959+650 (panel b), has been identified previously as an orphan flare. The other outliers can be interpreted in two different ways. One possibility is the existence of a mechanism producing additional gamma-rays, such as EC with seed photons from a region external to the jet, or hadronic processes involving accelerating protons interacting with jet photons. Another possibility is that some X-ray emissions are attenuated by intervening media on their way to Earth. Whatever the explanation, it should account for outliers across all the studied blazars and not be solely a characteristic of very high state fluxes. The exceptionally bright flare of PKS 2155-304 is entirely described by a quadratic correlation, consistent with single flares resulting from SSC mechanism and an increase of the electron injection \citep{singh2017time, singh2019multi}. 

\begin{figure}
    \centering
    \includegraphics[width=0.481\textwidth]{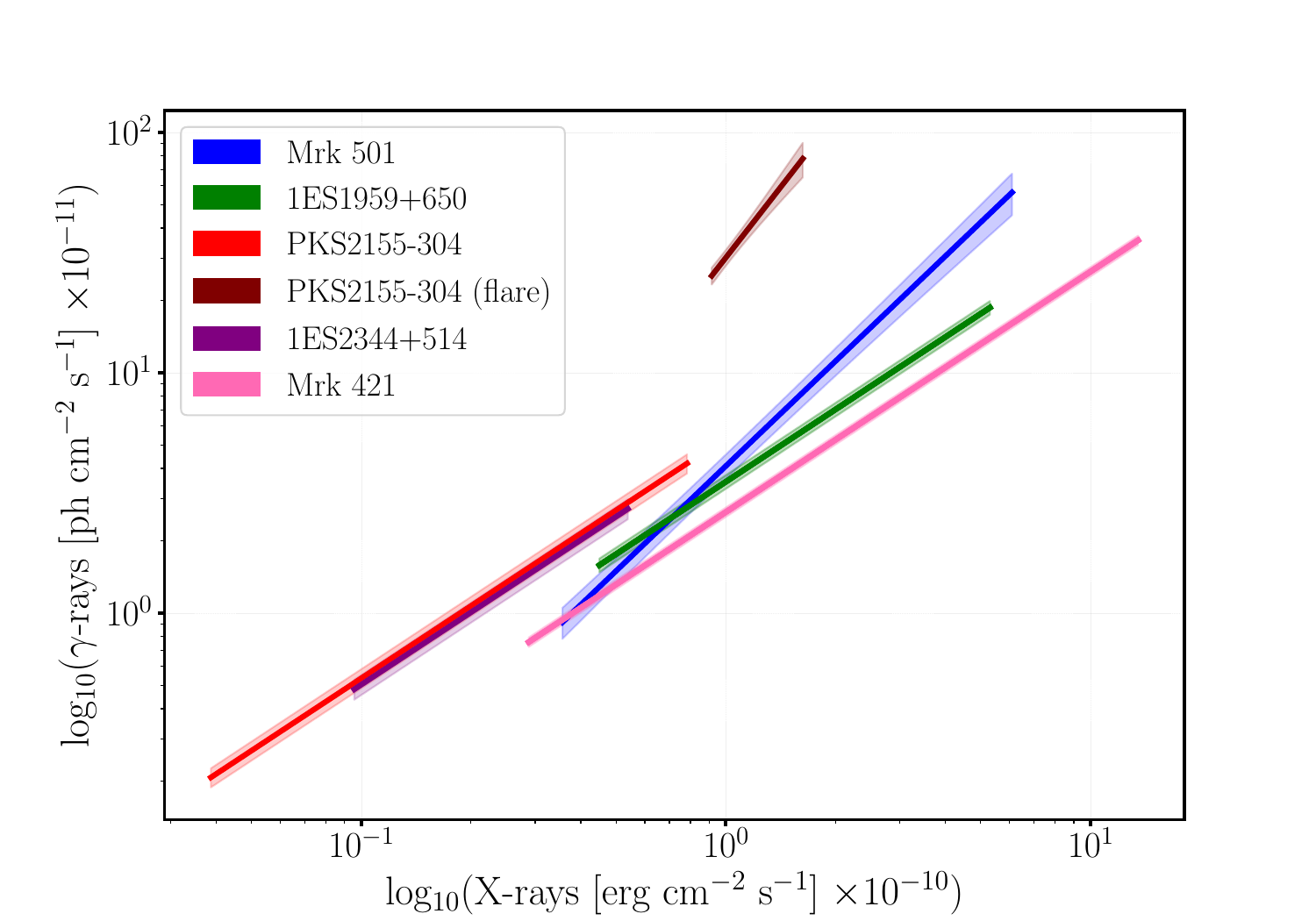}
    \caption{Comparison between the correlation models obtained for the five blazars in our sample and the exceptional flare of PKS 2155-304. The X-rays are plotted within the energy range of 2 to 10 keV and the gamma-rays with an energy threshold $>$ 300 GeV. The shaded region on the graph corresponds to the statistical error of each correlation model. This graph reveals that all five sources have a similar tendency, except for the steeper correlation model of Mrk 501 and the exceptional flare of PKS 2155-304 observed on 2006 July. PKS 2155-304 and 1ES 2344+514 exhibit a comparatively lower X-ray flux range when contrasted with Mkr 501, 1ES 1959+650, Mrk 421 and the exceptional flare of PKS 2155-304. Similarly, the latter ones, minus the flare, exhibit a similar gamma-ray flux output. Notably, the exceptional flare of PKS 2155-304 displays a gamma-ray flux range that is two orders of magnitude higher than its typical gamma-ray flux.
    }
    \label{fig:generalcorrelation}
\end{figure}

Figure \ref{fig:generalcorrelation} shows the preferred correlation models in a log-log plot for the four studied sources and Mrk 421, including the exceptionally bright flare of PKS 2155-304, with a gamma-ray energy threshold of $>$300 GeV and X-ray energy range from 2 to 10 keV. For Mrk 421, shown in pink, the correlation function is obtained with a change in the energy threshold (from $>$400 GeV to $>$300 GeV) using a power-law with a cutoff spectral model from a low-activity state \citep{Acciari2014APh....54....1A}. We observed that using a high-activity state model \citep{Accicari2011ApJ...738...25A} results in a very similar correlation. This line appears below the other correlations, which could be attributed to uncertainties from the energy threshold change and the conversion factor from Crab units to $\rm ph\ cm^{-2}s^{-1}$ \citep{Grube2008ICRC....2..691G}. 



The correlation observed during the exceptionally bright flare of PKS 2155-304 is notable for its quadratic behaviour, accompanied by gamma-ray fluxes significantly higher than those observed in other correlations for similar X-ray fluxes. While a quadratic correlation suggests that the emission mechanism responsible for the flare could be SSC, the extremely high gamma-ray fluxes raise the possibility of an additional mechanism contributing to this excess. This could involve another emission zone capable of producing very high gamma-ray fluxes, or an improbable scenario where X-ray attenuation increases during the flare. Importantly, all the other observations of PKS 2155-304 exhibit a linear correlation. 

Surprisingly, all studied sources show alignment with a general correlation, implying a common acceleration mechanism, with variations in the X-ray and gamma-ray output likely stemming from differences in individual magnetic field strengths. The correlation observed in Mrk 501 also follows this general trend, exhibiting a correlation stronger than linear but weaker than quadratic. However, Mrk 501 stands apart from the other sources due to its synchrotron peak shift up to two orders of magnitude higher during flares. Additionally, the X-ray flux increases to the point where the gamma-ray dominance occurs ($F_{\gamma} / F_{X} < 1 $). This relationship depends on factors such as magnetic field strength ($B$), Doppler factor ($\delta$), and the size of the emission region ($R$), where $F_{\gamma} / F_{X} \propto 1/ R^2 B^2 \delta^4$ \citep{tavecchio1998constraints}. The observed increase in both energy peak and flux range suggests a simultaneous acceleration of electrons and injection of particles within the emission region \citep{Acciari2020ApJS..247...16A}. Hence, the presence of a higher gamma-ray flux relative to expected compared to the corresponding X-ray flux might indicate the involvement of an additional radiation mechanism that exclusively generates gamma-rays (such as hadronic processes or an EC, etc). 


\begin{table*}
    \centering
    \caption{ \normalsize
Table including the correlation indexes obtained for each source  and the values of the AIC for each model. Columns 6, 7 and 8 show the Relative Likelihood between two AICs. In most cases, excluding Mrk 501, the simplest model—the linear one—was selected.}    
    \label{tab:aic}
\label{tab:fitresult}
\begin{tabular}{cccccccc}
\hline
\hline
Source & Correlation index  & AIC$_1$ & AIC$_{\rm free}$ & AIC$_2$ & RL$_{\rm free / 1}$ (\%) & RL$_{\rm free/ 2}$ (\%) & RL$_{\rm 1/ 2}$ (\%)  \\ \hline \hline 
Mkn 421   & 0.87 $\pm$ 0.08 & \textbf{116.82} & 116.69 & 181.48 & 94 & 8E-13 & 9E-13  \\
Mkn 501 & 1.45 $\pm$ 0.01 & 168.28 & \textbf{144.25} & 170.03 & 6E-4 & 2E-4 & 42 \\
1ES 1959+650 & 1.42 $\pm$ 0.22 & \textbf{372.60} & 370.71 & 374.35 & 39 & 16 & 42 \\
PKS 2155-304 & 0.54 $\pm$ 0.12 & \textbf{148.54} & 142.72 & 172.59 & 6 & 3E-5 & 6E-4 \\
PKS 2155-304 (flare) & 1.95 $\pm$ 0.29 & 96.17 & 90.21 & \textbf{86.57} & 5 & 16 & 0.82 \\
1ES 2344+514 & 1.25 $\pm$ 0.22 & \textbf{9.73} & 10.90 & 15.39 & 56 & 10 & 6  \\\hline \hline
\end{tabular}
\end{table*}


\begin{multicols}{2}
\end{multicols}
\begin{multicols}{2}
\end{multicols}


\section{Conclusions} \label{sec:conclusions}

In this study, we confirm the existence of a correlation between soft X-rays and TeV gamma-rays across multiple observational periods of observation in four HBL blazars Mrk 501, 1ES 1959+650, PKS 2155-304, and 1ES 2344+514. 
Most of our sources exhibit a correlation consistent with a linear model, except for Mrk 501, which shows a higher correlation index of 1.45. The steepness of the correlation in Mrk 501, along with observed outliers in several sources like in Mrk 421, and the presence of extreme flares like the one from PKS 2155-304, suggest an excess production of gamma rays. This could be indicative of a radiative mechanism more efficient than the standard SSC process in generating gamma-rays, which appears to be prevalent in many, if not all, blazars.

Future studies could extend this analysis to a larger sample of HBL blazars and include other types of blazars, such as IBL blazars, to explore correlations across a broader range of energies compared to those studied here.

Our analysis and interpretation of the results are limited due to the quasi-simultaneity of X and gamma-rays, highlighting the importance of a continuous monitoring of blazars across multiwavelenths. Water Cherenkow Detectors (WCD) such as the High-Altitude Water Cherenkov Observatory (HAWC), Large High Altitude Air Shower Observatory (LHAASO), and the upcoming Southern Wide-field Gamma-ray Observatory (SWGO) provide or will provide continuous monitoring of the TeV gamm-ray sky.  Similarly, X-ray satellites like Swift (XRT and BAT) contribute with crucial simultaneous observations.


\section*{Acknowledgements}

This work was supported by the UNAM-PAPIIT project number IG101323 and by project Gestiona I+D 02-2021 of Secretaria Nacional de Ciencia y Tecnología de Guatemala (SENACyT).

\section*{Data Availability}

The data used for this paper are all publicly available in the corresponding multiwavelenght works. The codes to implement the threshold changes and to model the correlations can be provided upon reasonable request.



\bibliographystyle{mnras}
\bibliography{bibliography} 







\bsp	
\label{lastpage}
\end{document}